\documentclass[journal]{IEEEtran}

\newcommand{\beq}{\begin{equation}}
\newcommand{\eeq}{\end{equation}}
\newcommand{\bsq}{\begin{subequations}}
\newcommand{\esq}{\end{subequations}}
\newcommand{\bq}{\begin{eqnarray}}
\newcommand{\eq}{\end{eqnarray}}
\newcommand{\bqn}{\begin{eqnarray*}}
\newcommand{\eqn}{\end{eqnarray*}}
\usepackage{bbm}
\usepackage[pdftex]{graphicx}
\graphicspath{{Figs/}}
\usepackage{enumerate}
\usepackage{enumitem} 
\usepackage[hidelinks]{hyperref}

\usepackage{mathptmx}       % selects Times Roman as basic font
\DeclareMathAlphabet{\mathcal}{OMS}{cmsy}{m}{n}
\usepackage{helvet}         % selects Helvetica as sans-serif font
\usepackage{courier}        % selects Courier as typewriter font
\usepackage{type1cm}        % activate if the above 3 fonts are
\usepackage{makeidx}         % allows index generation
\usepackage{graphicx}        % standard LaTeX graphics tool
\usepackage{url} 
\usepackage{multicol}        % used for the two-column index
\usepackage[bottom]{footmisc}% places footnotes at page bottom
\usepackage{ifthen}
\usepackage{subfigure}
\usepackage[usenames,dvipsnames]{color}

\makeindex             % used for the subject index
\usepackage{graphics} % for pdf, bitmapped graphics files
\usepackage{graphicx} % for pdf, bitmapped graphics files
\usepackage{epsfig} % for postscript graphics files
\usepackage{epstopdf}
\usepackage{mathptmx} % assumes new font selection scheme installed
\usepackage{times} % assumes new font selection scheme installed
\usepackage[cmex10]{amsmath}

\usepackage{mathtools}  % loads amsmath
\usepackage{amssymb}  % assumes amsmath package installed
\usepackage{amsthm}
\allowdisplaybreaks[4]

\usepackage{amsfonts}
\usepackage{cite}
\usepackage{verbatim}
\usepackage{comment}
\usepackage{algorithm}   %CZ commented out due to compiling error
\usepackage{algorithmic}
\usepackage{multirow}
\usepackage{cite}
\usepackage{stfloats}
\usepackage{supertabular}
\usepackage{longtable}

\DeclareGraphicsExtensions{.pdf,.png,.jpg,.eps,.ps}
\renewcommand{\arraystretch}{1.2}
\usepackage{arydshln}
\usepackage{multicol}
\usepackage{colortbl}
\newtheorem{myDef}{Definition}
\theoremstyle{definition}
\newtheorem{theorem}{Theorem}

\theoremstyle{definition}

\newtheorem{remark}{Remark}

\ifCLASSINFOpdf
\else
\fi

\usepackage{comment}
\usepackage{ifthen}
\newboolean{showcomments}
\setboolean{showcomments}{true}
\newcommand{\ychen}[1]{\ifthenelse{\boolean{showcomments}}
        { \textcolor{red}{YC: #1}}}
\newcommand{\tongxin}[1]{\ifthenelse{\boolean{showcomments}}
        { \textcolor{blue}{(#1)}}{}}

\usepackage{amsmath}
\allowdisplaybreaks[4]

\begin{document}
%\title{Chance-Constrained Multi-Timescale Operation of Integrated Heat-Power System with Shared Energy Storage}
\title{A Multi-timescale and Chance-Constrained Energy Dispatching Strategy of Integrated Heat-Power Community with Shared Hybrid Energy Storage}
\author{Wenyi Zhang,
Yue Chen,~\IEEEmembership{Member,~IEEE},
Rui Xie,
Yunjian Xu,~\IEEEmembership{Member,~IEEE}
\thanks{This research was supported in part by the General Research Fund (GRF) project 14200720 of the Hong Kong University Grants Committee, and the	National Natural Science Foundation of China (NSFC) Project 62073273. (Corresponding to Yunjian Xu and Yue Chen)

W. Zhang, Y. Chen, R. Xie and Y. Xu are with the Department of Mechanical and Automation Engineering, the Chinese University of Hong Kong, HKSAR, China (e-mail: wenyizhang@cuhk.edu.hk, yuechen@mae.cuhk.edu.hk, ruixie@cuhk.edu.hk, yunjianxu@mae.cuhk.edu.hk).}}

\markboth{Journal of \LaTeX\ Class Files,~Vol.~18, No.~9, September~2020}%
{How to Use the IEEEtran \LaTeX \ Templates}

\maketitle

\begin{abstract}
The community in the future may develop into an integrated heat-power system, which includes a high proportion of renewable energy, power generator units, heat generator units, and shared hybrid energy storage. In the integrated heat-power system with coupling heat-power generators and demands, the key challenges lie in the interaction between heat and power, the inherent uncertainty of renewable energy and consumers' demands, and the multi-timescale scheduling of heat and power. In this paper, we propose a game theoretic model of the integrated heat-power system. For the welfare-maximizing community operator, its energy dispatch strategy is under chance constraints, where the day-ahead scheduling determines the scheduled energy dispatching strategies, and the real-time dispatch considers the adjustment of generators. For utility-maximizing consumers, their demands are sensitive to the preference parameters. Taking into account the uncertainty in both renewable energy and consumer demand, we prove the existence and uniqueness of the Stackelberg game equilibrium and develop a fixed point algorithm to find the market equilibrium between the community operator and community consumers. Numerical simulations on integrated heat-power system validate the effectiveness of the proposed multi-timescale integrated heat and power model.
\end{abstract}


\section*{Nomenclature}
\addcontentsline{toc}{section}{Nomenclature}

\subsection{Acronyms}
\begin{IEEEdescription}[\IEEEusemathlabelsep\IEEEsetlabelwidth{sssssssss}]
	\item[SOC] State of charge
        \item[ES]  Energy storage
	\item[SES] Shared electricity storage
	\item[SHS] Shared heat storage
	\item[SHES] Shared hybrid energy storage
	\item[GMM] Gaussian mixture model
        \item[DAS] Day-ahead scheduling
        \item[RTD] Real-time dispatching
	\item[UoG] Uncertainty of generators
	\item[UoC] Uncertainty of consumers
	\item[CHP] Combined heat-power
	\item[KKT] Karush-Kuhn-Tucker
	\item[IHP] Integrated heat-power
        \item[TLOU] Time-and-level-of-use price
\end{IEEEdescription}

\subsection{Parameters}
\begin{IEEEdescription}[\IEEEusemathlabelsep\IEEEsetlabelwidth{sssssssss}]
	\item[${\overline P _i}$, $\underline P _i$] Upper and lower boundaries of the generators
	\item[$R{u_i}$, $R{d_i}$] Upper and lower boundaries of ramping rate of generator $i$	
	\item[${\eta _{ph}}$, ${\eta _{gh}}$]  Efficiency of CHP and heat-power units. 
	\item[${\overline P _{ut}}$, $	{\underline P _{ut}}$] Upper and lower boundaries of trading power to CHP units. 
	\item[${\overline H _{gt} }$, ${\underline H _{gt}}$] Upper and lower boundaries of the trading gas. 	
	\item[${P_c}$, ${P_d}$] Upper boundaries of the charging and discharging rate of SES.
	\item[${Q_s}$, ${Q_{hs}}$]  Total capacity of the SES and SHS. 
	\item[${S_u}$, ${S_l}$] Upper and lower boundaries of the SOC. 
	\item[${\eta _c}$, ${\eta _d}$]  Charging and discharging efficiencies of the SES.  
	\item[${H_c}$, ${H_d}$] Upper boundaries of the SHS charging and discharging rates.
	\item[${S_{hu}}$, ${S_{hl}}$] Upper and lower boundaries of the SHS SOC.
	\item[${\eta _{hc}}$, ${\eta _{hd}}$] Charging and discharging efficiencies of SES.
	\item[${\alpha ^{L + }},{\alpha ^{L - }}$] Maximal tolerable risk of the upward and downward power flow.
	\item[${\bar P}^L$] Maximal transmission capacity.
	\item[$\alpha$] Preference parameter of the utility function.
	\item[$\lambda _{pt}$, $\lambda _{ht}$] Power and heat prices.
	\item[$\bar O$]  Upper bound of the user utility.
	\item[$\lambda _h^t$]  Gas price at time $t$.
	\item[${K_1}$, ${K_2}$] Curtailment penalty coefficients.
	\item[$C_1$, $C_2$] Maximal capacity of the wind and solar energy sources. 
	\item[$\Phi _{w,forecast}^t$] Expected forecast wind power. 
	\item[$\Upsilon _{s,forecast}^t$] Expected forecast solar power.
	\item[$ {\gamma _i}$] Participation factor of affine control
	\item[${\alpha _{up}}$, ${\alpha _{dr}}$] Maximal tolerable risk of insufficient upward  and  downward reserve capacity.		
	\item[${{\bar P}^L}$] Maximal transmission capacity.
	\item[$c_i$] Cost coefficient of generator $i$.
	\item[$r_{ou}$, $x_{ou}$]  Resistance and reactance of line $ou$.	
	\item[${T_a}$]  The ambient temperature.
	\item[${{H_{nt}}}$] Heat demands of the consumer $n$ at time $t$.
	\item[${c_p}$] Specific heat capacity of water.  
	\item[$\lambda _0$] Heat leakage loss coefficient. 
	\item[${L_0}$] Pipeline length.
	\item[$\dot m$] Mass flow rate.
	\item[${e_{h,ini}}$, ${e_{ini}}$] Initial states of SHS and SES.
	\item[$Q\left( {\varepsilon |p} \right)$] 	p-quantile of random variable $\varepsilon$.
\end{IEEEdescription}

\subsection{Decision Variables}
\begin{IEEEdescription}[\IEEEusemathlabelsep\IEEEsetlabelwidth{ssssssss}]
	\item[$p_i^t $] Output of generator $i$ at time $t$	
	\item[${h_{ut}}$] Transformed heat from the CHP units. 
	\item[$p_{ut}$] Power transformed to heat at time $t$. 
	\item[$h_{gt}$] Transformed heat from heat-power units. 
	\item[$h_{gas,t}$] Purchased gas power at time $t$. 
	\item[${{{e}}_t}$, ${{{e}}_{ht}} $]  SOC of SES and SHS.  
	\item[$p_{st}^c$, $p_{st}^d$] Charging and discharging power of SES.
	\item[$h_{st}^c$, $h_{st}^d$] Charging and discharging power of SHS.
	\item[$\bar \phi _w^t$, $ \bar \zeta _s^t$] Allowable upper limit of wind and solar generation.  
	\item[$\phi _{sch}^t $, $\zeta _{sch}^t$] Scheduled power generation of wind and solar power. 
	\item[$\widetilde \phi _w^t$, $\widetilde \zeta _s^t$] Actual power output of wind and solar generator.
	\item[$\widetilde \phi _{w,av}^t$] Available power output of wind generator.
	\item[$\widetilde \zeta _{s,av}^t$] Available power output of solar generator.
	\item[$l_{nt}^d$] Power demands of the consumer $n$ at time $t$.
 	\item[$\widetilde p_i^t$] Real time generation power.
  	\item[$q_i^t$] Unit reactive power output of generator $i$.
	\item[$h_{nt}^d$] Heat demands of the consumer $n$ at time $t$.	
	\item[$v_{ot}$, $v_{ut}$]  Voltage magnitude square at original bus $o$ and  terminal bus $u$.
	\item[$ P_{out}$, $Q_{out}$] Active and reactive power flow on  transmission line $ou$.	
 	\item[$P_{ut}^{in}$, $Q_{ut}^{in}$] Active and reactive power injection at bus $u$.	
	\item[$l_{ou}$] Current square of line $ou$.
	\item[$p_i^{tl}$, $q_i^{tl}$] Active and reactive power demands at bus $i$.
	\item[$T_{{v_1}}$, ${T_{{v_2}}}$]  Supply and outlet water temperature of pipeline.
\end{IEEEdescription}

\section{Introduction}
\IEEEPARstart{N}{OWADAYS}, with the rapid increase of the global population and advances in civilization, the energy demand is in exponential growth  \cite{ola2022}. The consumption of fossil fuels also rises accordingly. The combustion of these fossil fuels results in the emission of a substantial quantity of carbon dioxide and other greenhouse gases into the atmosphere, thereby increasing the risks of global greenhouse effect and climate change. \cite{liu2023energy}. China will strive to achieve carbon neutrality by 2060 by reducing carbon emissions, increasing carbon absorption and storage, and making carbon emissions zero. This will be achieved through measures such as promoting renewable energy development. \cite{liang2022}. However, the intermittency and uncertainty of renewable energy remain to be solved\cite{Nosair2015}.

With the development of electrical heating devices \cite{li2015}, the electricity market is coupled with district heat systems through the power to heat turbines \cite{Doroti2020}. However, The increasing development of renewable energy sources requires more flexible technologies to be applied in heat-power systems \cite{Bagherian2020}. 

Energy storage (including both electricity and heat storage) is an essential way to enhance the resilience of the IHP system, and to balance the uncertainty of renewable energy and reducing operation costs \cite{Niu2019}. The traditional approach of individual distributed ES is deploying individual energy storage units for consumers, allowing them to store electricity when needed and supply it when required \cite{Koohi2021}. Although the investment and operation costs of electricity storage \cite{Rotella2021} and heat storage \cite{Borri2021} have been decreasing recently, they are still high for grid-scale deployment.  In practice, innovating business models can lower the cost of energy storage, for example, by leveraging the sharing economy, establishing partnerships, or innovating pricing mechanisms to reduce costs and provide more flexible energy storage solutions. \cite{Baumgarte2020}. 

Shared energy storage enables cost reduction \cite{dai2021}, optimized resource utilization \cite{Walker2021}, and flexible energy storage \cite{Zhang2020}, making it an increasingly prominent and adopted innovative business model. Currently, ownership-based classification allows shared energy storage to be categorized into four distinct types: i) private ES, in which individual consumers own their own energy storage systems and participate in energy trading to share energy resources \cite{Kong2020,Zhang2022}, ii) the interconnected ES, in which the ES units owned by different consumers are connected, allowing them to share the capacity of their respective systems \cite{Tusharn2016}, iii) the common ES, in which the SES unit is installed and managed collectively by all consumers, providing a centralized storage solution \cite{Zhu2021}, and iv) the independent ES, where SES systems are managed by an independent operator, who oversees the operations and maintenance of the storage infrastructure \cite{Szab2020,Mediwaththe2020}. In this paper, we consider the SHES, which is a joint sharing of electricity and heat storage in a community and belongs to the community.

Although extensive research has been conducted on the operation of IHP systems \cite{Li2021}, it is important to highlight that challenges still exist regarding the scheduling and dispatching of these systems, particularly in the presence of uncertainties. Compared with \cite{Karami2017}, which only takes the demand uncertainties of the electric load into account, our model incorporates the uncertainty of wind and solar generation. 

In Table I, we compare the proposed model with existing ones on service pricing of shared energy storage \cite{Zhang2020,fang2016,de2018}, multi-timescale scheduling of microgrids \cite{bao2015}, and operation of heat-power system \cite{chen2018,li2022enhancing,Shang20171,Bao2022,Wei2020}. Different from \cite{Zhang2020,fang2016,de2018}, in this paper, heat demands and power demands are coupled through the Cobb-Douglas utility function, and the supply of heat and power is coupled through the CHP turbines. The shared hybrid energy storage comprises shared electricity and heat storage.

\begin{table}[t]
	\renewcommand{\arraystretch}{1.3}
	\renewcommand{\tabcolsep}{1em}
	\scriptsize
	\centering
	\caption{Comparision with existing models}
	\label{tab:comparison-1}
	\begin{tabular}{ccccccc}
		\hline  {Topic} & {Heat-power} & {Multi-timescale} & {SES} & {STS} & {UoG} & {UoC} \\
		\hline
	{\cite{Zhang2020}} &  &  & $\checkmark$ & {} & $\checkmark$ & $\checkmark$\\
	{\cite{fang2016}} &$\checkmark$  &  & $\checkmark$ &  & & \\
	{\cite{de2018}} &$\checkmark$  &  & $\checkmark$ &  & & \\
	{\cite{bao2015}} &  & $\checkmark$ &  &  &$\checkmark$ & \\
	{\cite{chen2018}} & $\checkmark$ &  &  &  & & \\
	{\cite{li2022enhancing}} &$\checkmark$  &  & $\checkmark$  & $\checkmark$ & & \\
	{\cite{Shang20171}} &$\checkmark$  &  &  & $\checkmark$ & & \\
	{\cite{Bao2022}} &$\checkmark$  &  &  &  &$\checkmark$ & \\
	{\cite{Wei2020}} &$\checkmark$  &  &  &  &$\checkmark$ & \\
	{This work} & $\checkmark$ & $\checkmark$ & $\checkmark$ & $\checkmark$ & $\checkmark$&$\checkmark$ \\
		\hline
	\end{tabular}
\end{table}

This research is focused on the chance-constrained multi-timescale operation of the IHP system with SHES considering uncertainties. Compared with the previous heat and power network scheduling research, the main contribution of this work is two-fold.

1) From a modeling perspective: We propose a Stackelberg game model of the integrated heat-power system where the community operator acts as a leader and consumers act as followers in a two-stage chance-constrained model. Different from the conventional heat-power system, the operator considers the multi-timescale operation of multi-energy sources, and the consumers aim at maximizing utility by choosing their power and heat loads according to the energy price and their preference. What is more, SHES is incorporated into the integrated heat-power system model, which includes both shared electricity and heat storage. The interaction among heat and power plants is explicitly modelled.

The proposed model explicitly incorporates the uncertainties of renewable energy, which lie in the difference between the DAS and the real-time output of renewable energy. The uncertainty of the preference-sensitive consumers' demands is incorporated by the Cobb-Douglous function, which models the utility maximization of heat and power users.  As such, the proposed model captures the key features of the multi-timescale operation for the integrated heat-power system with shared energy storage under uncertainties.

2) From a solution standpoint: We propose a multi-timescale scheduling scheme and develop a dispatching strategy under chance constraints. In DAS, the GMM model is employed to represent the uncertainty surrounding wind and solar power generation. In RTD, the uncertainties of the wind and solar generations are balanced by generators. We establish the existence and uniqueness of the load equilibrium and propose a fixed-point algorithm to compute the load equilibrium. 

The remainder of this paper is organized as follows. Section II describes the model formulation of an integrated heat-power system with SHES. The solutions are proposed in Section III. Numerical simulations are carried out in Section V. In Section VI, we make some brief concluding remarks.

\section{Component Modeling}

The mathematical formulation of the proposed heat-power system with SHES will be presented in this section.

\subsection{Structure}
A representative structure of the heat-power system with SHES is shown in Fig. 1, including power plants, heat plants, SHES, and consumers. The power plants, heat plants and SHES are owned and shared by the consumers in the community. Market clearing and coordination with the community operator are handled by a third-party organizer.

\begin{figure}[!t]
	\centering
	\includegraphics[width=3in]{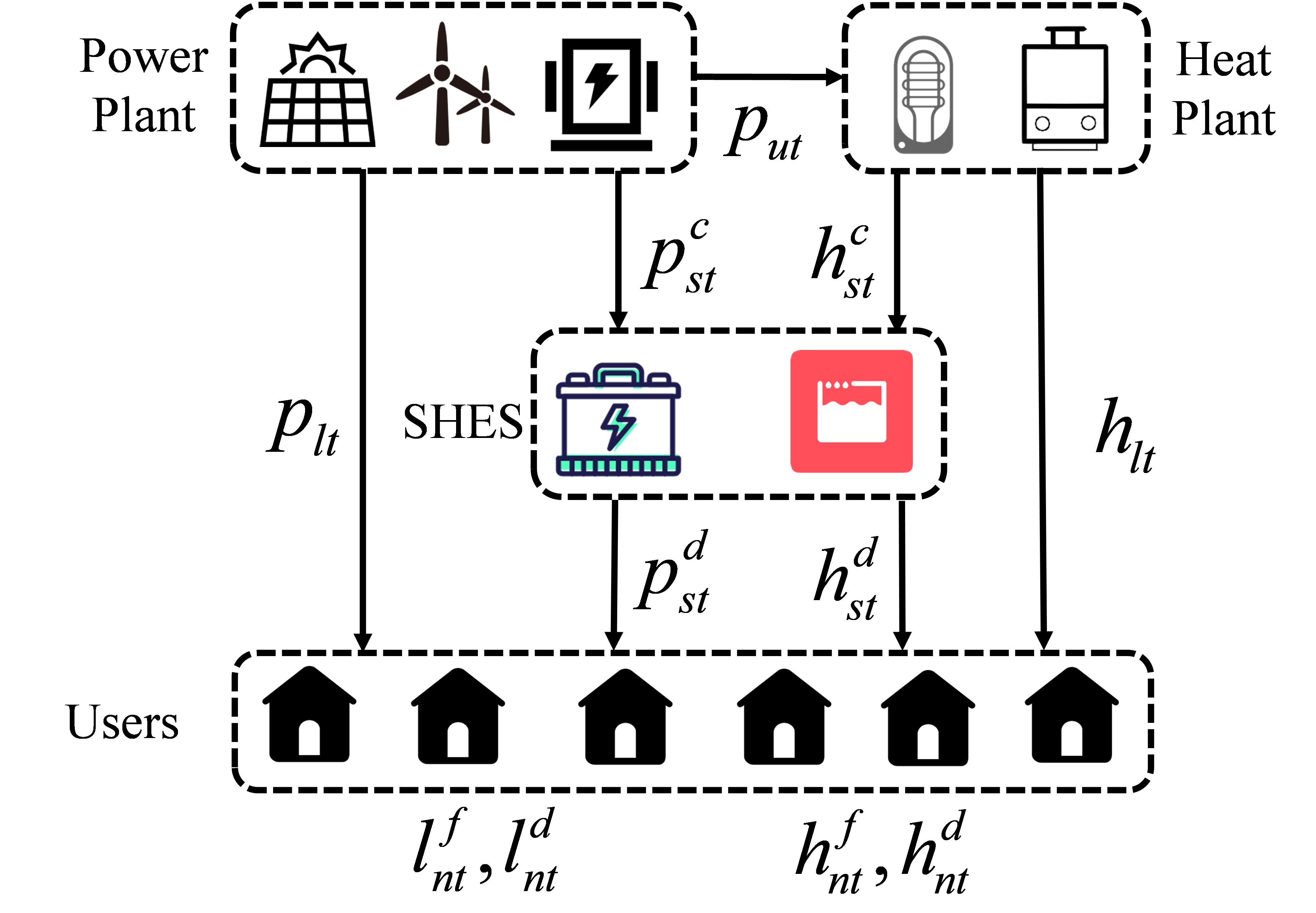}
	\caption{The structure of the heat-power system with SHES.}
	\label{structure1}
\end{figure}

\subsection{Power Plants}
Power plants have wind turbines, solar photovoltaics, and backup generators to support the power supply. When renewable energy generation can not meet consumers' demands, the backup generators can meet the demands of the consumers.  We consider the curtailment of renewable energy, and the renewable energy constraints can be formulated as follows:
\begin{equation}
0 \leqslant \phi _{sch}^t \leqslant \Phi _{w,forecast}^t,  \label{eq1}
\end{equation}
\begin{equation}
0 \leqslant \zeta _{sch}^t \leqslant \Upsilon _{s,forecast}^t,  \label{eq2}
\end{equation}
\begin{equation}
\widetilde \phi _w^t = \left\{ {\begin{array}{*{20}{l}}
	{\bar \phi _w^t\left( {if\widetilde \phi _{w,av}^t \geqslant \bar \phi _w^t} \right)} \\ 
	{\widetilde \phi _{w,av}^t\left( {if\widetilde \phi _{w,av}^t < \bar \phi _w^t} \right)} 
	\end{array}} \right.,\label{eq3}
\end{equation}
\begin{equation}
\widetilde \zeta _s^t = \left\{ {\begin{array}{*{20}{l}}
	{\bar \zeta _s^t\left( {if\widetilde \zeta _{s,av}^t \geqslant \bar \zeta _s^t} \right)} \\ 
	{\widetilde \zeta _{s,av}^t\left( {if\widetilde \zeta _{s,av}^t < \bar \zeta _s^t} \right)} 
	\end{array}} \right.,\label{eq4}
\end{equation}
where constraints (\ref{eq1}) and (\ref{eq2}) stipulate the scheduled wind and solar power generation should be lower than the forecast value. Constraint (\ref{eq3}) and (\ref{eq4}) stipulate the actual wind and solar power output should be lower than the allowable upper bound $\bar \phi _w^t$ and $\bar \zeta _s^t $.

The operation constraints of the generators are as follows:

\begin{equation}
{\underline P _i} \leqslant p_i^t \leqslant {\overline P _i},\label{eq5}
\end{equation}
\begin{equation}
p_i^t - p_i^{t - 1} \leqslant R{u_i}, \quad p_i^{t - 1} - p_i^t \leqslant R{d_i},\label{eq6}
\end{equation}
where constraint  (\ref{eq5}) requires the upper and lower boundaries of the generator $i$. Constraint (\ref{eq6}) determines the upper and lower boundary of the ramping rate of generator $i$.

\subsection{Heat Plants}
Heat plants include power-heat and gas-heat turbines. The power-heat turbine can generate heat by electricity. The gas-heat turbine can generate heat by gas. The heat plants would choose the gas source or power source according to the total cost. The operation constraints of heat plants can be described as follows:
\begin{equation}
{h_{ut}} = {\eta _{ph}}{p_{ut}},\quad {h_{gt}} = {\eta _{gh}}{h_{gas,t}},\label{eq7}
\end{equation}
\begin{equation}
	{\underline P _{ut}} \leqslant {p_{ut}} \leqslant {\overline P _{ut}},\quad {\underline H _{gt}} \leqslant {h_{gas,t}} \leqslant {\overline H _{gt}},\label{eq8}
\end{equation}
where constraint  (\ref{eq7}) stipulates the transformed heat from CHP units from electricity and gas. Constraint  (\ref{eq8}) stipulates the upper and lower boundaries of the trading power to CHP units and the trading gas to heat-power units at time $t$.

\subsection{Shared Hybrid Energy Storage}

The SHES units, including SES and SHS, are owned by the community consumers. The SES can be charged or discharged according to the power price signal. The SHS can be charged or discharged according to the heat price signal. In the community, our emphasis lies on addressing the operational problem of SHES.  For the SHES, the operation constraints can be cast as follows:
\begin{equation}
	0 \le p_{st}^c \le {P_c}{Q_s},\quad 0 \le p_{st}^d \le {P_d}{Q_s},\label{eq9}
\end{equation}
\begin{equation}
	{S_l}{Q_s} \le {{{e}}_t} \le {S_u}{Q_s}\label{eq10},
\end{equation}
\begin{equation}
	{{{e}}_t} = {{{e}}_{t - 1}} + {\eta _c}p_{st}^c\Delta t - p_{st}^d\Delta t/{\eta _d}\label{eq11},
\end{equation}
\begin{equation}
	0 \le h_{st}^c \le {H_c}{Q_{hs}},\quad 	0 \le h_{st}^d \le {H_d}{Q_{hs}}\label{eq12},
\end{equation}
\begin{equation}
	{S_{hl}}{Q_{hs}} \le {{{e}}_{ht}} \le {S_{hu}}{Q_{hs}}\label{eq13},
\end{equation}
\begin{equation}
	{{{e}}_{ht}} = {{{e}}_{ht - 1}} + {\eta _{hc}}h_{st}^c\Delta t - h_{st}^d\Delta t/{\eta _{hd}}\label{eq14},
\end{equation}
where constraint  (\ref{eq9}) stipulates the upper and lower bound of the charging power $p_{st}^c$ and discharging power $p_{st}^d$. Constraint  (\ref{eq10}) stipulates the upper and lower bound of the SOC of the SES $e_t$.  Constraint  (\ref{eq11}) stipulates the relationship between $e_{t}$ and $e_{t-1}$. Constraint  (\ref{eq12}) stipulates the upper and lower bound of the charging heat $h_{st}^c$ and discharging heat $h_{st}^d$. Constraint  (\ref{eq13}) stipulates the upper and lower bound of the SOC of the SHS $e_{ht}$. Constraint  (\ref{eq14}) stipulates the relationship between ${{e}}_{ht}$ and $ {{{e}}_{ht - 1}}$.

\subsection{Time-and-Level-of-Use Price}
The pricing of electricity in this paper is that each consumer buys it based on both the time and level of usage \cite{Zhang2020,Zhang2022}.
\begin{equation}
{\lambda _{pt}} = \lambda _{pt}^0 + kl_{nt}^d,\quad {\lambda _{ht}} = \lambda _{ht}^0 + kh_{nt}^d,\label{eq99}
\end{equation}
where $\lambda _{pt}^0$ and $\lambda _{ht}^0$ depend on time.  $kl_{nt}^d$ and $kh_{nt}^d$ depend on power and heat demands.

\subsection{Consumers}
In the community, each consumer aims to maximize their utility function. The preference parameters and market trading price can affect the demands of the consumers. The consumers' problem can be characterized as the Cobb-Douglas function \cite{ChenY2018,Chenavaz2022} as follows:
\begin{equation}
	{{{\max }_{{l_{nt}^d},{h_{nt}^d}}}U\left( {l_{nt}^d,h_{nt}^d} \right) = {{\left( {l_{nt}^d} \right)}^\alpha }{{\left( {h_{nt}^d} \right)}^{1 - \alpha }}},\label{eq15}
\end{equation}
\begin{equation}
	{s.t.  \quad \lambda _{pt}l_{nt}^d + \lambda _{ht}h_{nt}^d \le \bar O},\label{eq16}
\end{equation}
where constraint (\ref{eq16}) stipulates the upper bound of the utility. The monotonicity and concavity of the  Cobb-Douglas function are proven in the Appendix A \textcolor{red}{of [xx]}.

The optimal solution to problem (\ref{eq15})-(\ref{eq16}) can be obtained as follows:
\begin{equation}
\frac{{\lambda _{pt}{l_{nt}^d}}}{\alpha } = \frac{{\lambda _{ht}{h_{nt}^d}}}{{1 - \alpha }} = \bar O.\label{eq17}
\end{equation}

\section{Multi timescale and multi-energy scheduling}
 
The framework of the proposed multi-timescale dispatching model is shown in Fig. 2. The energy dispatching of an integrated heat-power community with SHES includes a DAS model and a real-time dispatching model. In the DAS model, the schedules of generators, heat demands, electricity demands and SHES of the next day are determined. In RTD, the time scale of heat energy dispatching is 1 hour, and the time scale of electrical energy dispatching is five minutes. Once a RTD cycle is completed, the time window advances by a one-time scale. The mathematical formulation of the proposed IHP system with SHES will be presented.

\begin{figure}[!t]
	\centering
	\includegraphics[width=3.2in]{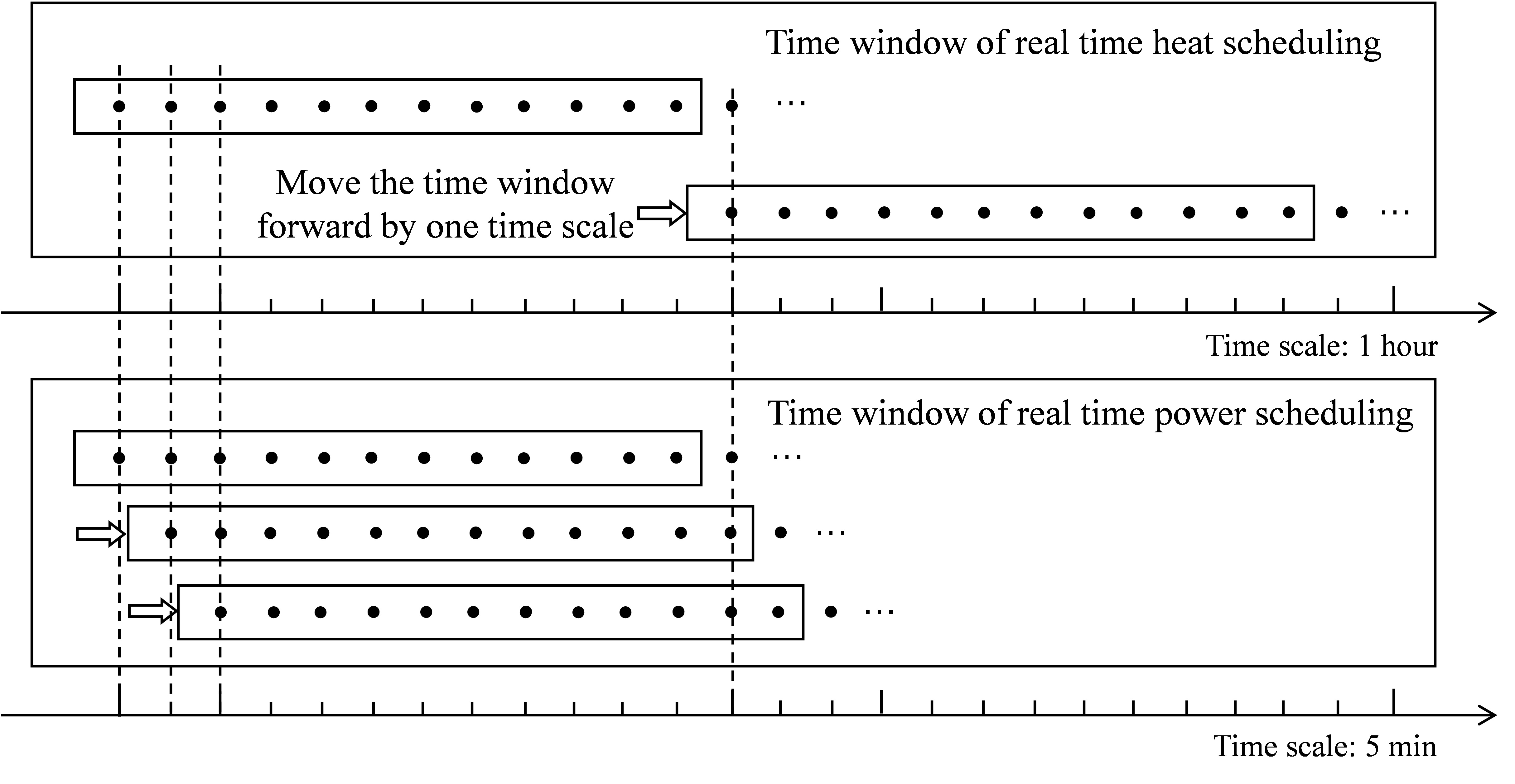}
	\caption{The solution structure of the heat-power system with SHES.}
	\label{multitime}
\end{figure}

\subsection{Day-Ahead Scheduling Model}

The IHP system with the SES model in DAS model can be cast as follows:

\subsubsection{Objective function}
The community operators aim to minimize the total costs, which is the sum of power generation costs, heating costs, and curtailment penalty costs. The objective function can be formulated as follows:
\begin{equation}
\begin{gathered}
\textbf{P1:}\min \sum\limits_{t = 1}^T {\sum\limits_{i = 1}^{{N_G}} {\left( {{a_i}{{\left( {p_{_i}^t} \right)}^2} + {b_i}p_{_i}^t + {c_i}} \right)} }  + \sum\limits_{t = 1}^T {{h_{gas,t}}\lambda _h^t}  \hfill \\
+ \sum\limits_{t = 1}^T {\left( {W_1^t\left( {\bar \phi _w^t} \right) + W_2^t\left( {\bar \zeta _s^t} \right)} \right)}  \hfill \\ 
\end{gathered} ,   \label{eq18}
\end{equation}
where ${N_G}$ stipulates the number of generators. ${a_i}$, ${b_i}$ and ${c_i}$ are generation constants. $p_{_i}^t$ stipulates the power output of generator $i$. $h_{gas,t}$ stipulates the purchased gas at time $t$. $\lambda _h^t$ stipulates the gas price at time $t$. $W_1^t\left( {{{\bar \phi }_t}} \right)$ stipulates the curtailment penalty of wind power. $W_2^t\left( {{{\bar \zeta }_t}} \right)$ stipulates the curtailment penalty of solar power.

The curtailment penalties of wind power and solar power are given by 
\begin{equation}
W_1^t\left( {\bar \phi _w^t} \right) = {K_1}\int_{\bar \phi _w^t}^{{C_1}} {\left( {\phi  - \bar \phi _w^t} \right)PD{F_t}\left( \phi  \right)d\phi } ,\label{eq19}
\end{equation}
\begin{equation}
W_2^t\left( {\bar \zeta _s^t} \right) = {K_2}\int_{\bar \zeta _s^t}^{{C_2}} {\left( {\zeta  - \bar \zeta _s^t} \right)PD{F_t}\left( \zeta  \right)d\zeta } ,\label{eq20}
\end{equation}
where ${K_1}$ and ${K_2}$ are the constants.

\subsubsection{Affine control and reserve constraints}
The real-time power output of generator $i$ can be regulated by the affine control. The affine control method is adopted by the real-world AGC system. The reserve adequacy should be under high probability. The affine control and reserve constraints can be formulated as follows:
\begin{equation}
\widetilde p_i^t = p_i^t - {\gamma _i}\left( {\widetilde \phi _w^t - \phi _{w,sch}^t + \widetilde \zeta _s^t - \zeta _{s,sch}^t} \right),\label{eq21}
\end{equation}
\begin{equation}
	\Pr \left[ {\widetilde p_i^t \leqslant {{\overline P }_i}} \right] \geqslant 1 - {\alpha _{up}},\label{eq22}
\end{equation}
\begin{equation}
	\Pr \left[ {\widetilde p_i^t \geqslant {{\underline P }_i}} \right] \geqslant 1 - {\alpha _{dr}},\label{eq23}
\end{equation}
where constraint (\ref{eq21}) defines the real-time generation power $p_i^t$. Constraint (\ref{eq22}) and (\ref{eq23}) specify the chance constraints of the upward and downward reserve capacity.

\subsubsection{Transmission constraints}
In order to maintain the reliability of the power system, it is necessary for the power flow on the transmission line to remain below the maximum capacity for power transmission. The risk of overloading should be under low probability. The transmission constraints are given by
\begin{equation}
	\Pr \left[ {\sum\limits_{i = 1}^{{N_G}} {F_i^L\widetilde p_i^t}  + F_w^L\widetilde \phi _w^t + F_s^L\widetilde \zeta _s^t + \sum\limits_{n = 1}^{{N_D}} {F_n^L{D_{nt}}}  \leqslant {{\bar P}^L}} \right] \geqslant 1 - {\alpha ^{L + }},\label{eq24}
\end{equation}
\begin{equation}
	\Pr \left[ {\sum\limits_{i = 1}^{{N_G}} {F_i^L\widetilde p_i^t}  + F_w^L\widetilde \phi _w^t + F_s^L\widetilde \zeta _s^t + \sum\limits_{n = 1}^{{N_D}} {F_n^L{D_{nt}}}  \geqslant  - {{\bar P}^L}} \right] \geqslant 1 - {\alpha ^{L - }},\label{eq25}
\end{equation}
where $F_i^L$,$F_w^L$,$F_s^L$,$F_n^L$ are Power transfer distribution factors. Constraints (\ref{eq24}) and  (\ref{eq25}) stipulate the chance constraints of the power flow of the transmission line.

\subsubsection{Power flow constraints}
The power flow constraints of the IHP system are as follows:
\begin{equation}
{v_{ot}} - {v_{ut}} = 2\left( {{r_{ou}}{P_{out}} + {x_{ou}}{Q_{out}}} \right) - \left( {r_{ou}^2 + x_{ou}^2} \right){l_{out}},\label{eq26}
\end{equation}
\begin{equation}
P_{out}^2 + Q_{out}^2 = {l_{out}}{v_{ot}},\label{eq27}
\end{equation}
\begin{equation}
{P_{ut}^{in}} = \phi _{w,sch}^t + \zeta _{s,sch}^t + \sum\limits_{i = 1}^{{N_G}} {p_i^t}  + p_{st}^d - \sum\limits_{n = 1}^{{N_D}} {{l_{nt}^d}}  + p_{st}^c + {p_{ut}},\label{eq28}
\end{equation}
\begin{equation}
{Q_{ut}^{in}} = \sum\limits_{y:u \to y} {{Q_{uyt}}}  - \left( {{Q_{out}} - {l_{out}}{x_{ou}}} \right),\label{eq29}
\end{equation}
\begin{equation}
{P_{ut}^{in}} = p_i^t - p_i^{tl}, \quad{Q_{ut}^{in}} = q_i^t - q_i^{tl},\label{eq30}
\end{equation}
\begin{equation}
{\underline P _{ou}} \leqslant {P_{out}} \leqslant {{\bar P}_{ou}}, \quad  {Q_i} \leqslant q_i^t \leqslant {{\bar Q}_i},\label{eq31}
\end{equation}
\begin{equation}
{\underline v _i} \leqslant {v_{it}} \leqslant {\bar v_i},    \quad {l_{out}} \geqslant 0,\label{eq32}
\end{equation}
where constraint (\ref{eq26}) corresponds to the voltage balance. Constraint (\ref{eq27}) corresponds to the power balance of active and reactive power. Constraint (\ref{eq28}) limits the power balance of active power. Constraint (\ref{eq29}) limits the power balance of reactive power. Constraint (\ref{eq30}) stipulates the definition of injective active and reactive power. Constraint (\ref{eq31}) limits the bounds of the nodal voltage and the active power of line $ou$.  Constraint (\ref{eq32}) limits the bounds of the current of line $ou$ and the nodal reactive power.

\subsubsection{Heat network constraints}
The heat network constraints are as follows:
\begin{equation}
{\eta _{ph}}{p_{ut}} + {\eta _{gh}}{h_{gas,t}} + h_{st}^d = \sum\limits_{n = 1}^{{N_H}} {{h_{nt}^d}}  + h_{st}^c,\label{eq33}
\end{equation}
\begin{equation}
	{T_{{v_2}}} = \left( {{T_{{v_1}}} - {T_a}} \right){e^{ - \frac{{{\lambda _0}{L_0}}}{{{c_p}\dot m}}}} + {T_a},\label{eq34}
\end{equation}
\begin{equation}
	{\eta _{eh}}{p_u} + {h_k} = {c_p}\dot m\left( {{T_{vS}} - {T_{vR}}} \right),\quad v \in {M_p},\label{eq35}
\end{equation}
\begin{equation}
	h_v^{tl} = {c_p}\dot m\left( {{T_{vS}} - {T_{vR}}} \right),\quad v \in {M_L},\label{eq36}
\end{equation}
\begin{equation}
	{T_v} = \frac{{\sum\limits_{\kappa  \in {S_v}} {{{\dot m}_\kappa }} {T_\kappa }}}{{\sum\limits_{\kappa  \in {S_v}} {{{\dot m}_\kappa }} }},\quad v \in {M_{\text{m}}},\label{eq37}
\end{equation}
where constraint (\ref{eq33})  stipulates the heat balance. Constraint (\ref{eq34}) stipulates the temperature drop. Constraints (\ref{eq35})-(\ref{eq36}) stipulate the heat demands and the heat generation. Constraint (\ref{eq37}) stipulates the temperature of the mixed fluid.

\subsubsection{Component operation constraints}
The operation constraints of the power plants, heat plants and SHES are shown in (\ref{eq1})-(\ref{eq14}). Besides, the operation constraints of the SHES are shown as follows:
\begin{equation}
{e_{h0}} = {e_{h,ini}},\quad  {e_0} = {e_{ini}},\label{eq38}
\end{equation}
\begin{equation}
{e_{hT}} = {e_{h0}},\quad {e_T} = {e_0},\label{eq39}
\end{equation}
where ${e_{h,ini}}$ is the initial state of SHS. ${e_{ini}}$ is the initial state of SES. Constraint  (\ref{eq38})  stipulates the initial SOC of SHES. Constraint (\ref{eq39}) requires that the initial SOC is the same as the terminal SOC of SHES.

\begin{remark}
The curtailment of wind and solar power are decision variables in the DAS stage and given parameters in the RTD stage. In the operation service optimization model of the community operator (\ref{eq18})-(\ref{eq39}), the optimal generation schedule and the curtailment of renewable energy are obtained simultaneously.
\end{remark}

\subsection{Real-time  Dispatching Model}
In RTD, the objective of real-time electricity dispatching is to minimize the cost of backup adjustment under the  curtailment  of wind generation in DAS.  The real-time electricity dispatching optimization model is given as follows:
\begin{equation}
\textbf{P2:}\mathop {\min }\limits_{\widetilde p_i^t,p_i^t} \sum\limits_{i = 1}^{{N_G}} {c_i^{pen}\left( {\widetilde p_i^t - p_i^t} \right)}, \label{eq40}
\end{equation}
\begin{equation}
s.t.{\text{ }}\sum\limits_{i = 1}^{{N_G}} {\widetilde p_i^t}  = \sum\limits_{i = 1}^{{N_G}} {p_i^t}  - \left( {\widetilde \phi _w^t - \phi _{w,sch}^t + \widetilde \zeta _s^t - \zeta _{s,sch}^t} \right),\label{eq41}
\end{equation}	
\begin{equation}
(\ref{eq1})-(\ref{eq14}),\quad (\ref{eq21}),\quad (\ref{eq26})-(\ref{eq39}) ,\label{eq42}
\end{equation}
where (\ref{eq41}) represents the power balance constraint.

\begin{remark}
The real-time redispatch of conventional generator $\widetilde p_i^t$ is carried out in the RTD stage, given the scheduled output of generator $p_i^t$, the scheduled wind generation  $\phi _{w,sch}^t $, and solar generation $\zeta _{s,sch}^t$ in the DAS stage. The RTD model incorporates real-time wind and solar generation uncertainties.
\end{remark}

\section{Soltuion}
In this section, the solution structure of the bi-level IHP model is introduced, as shown in Fig. 3.
\begin{figure}[!t]
	\centering
	\includegraphics[width=3.5in]{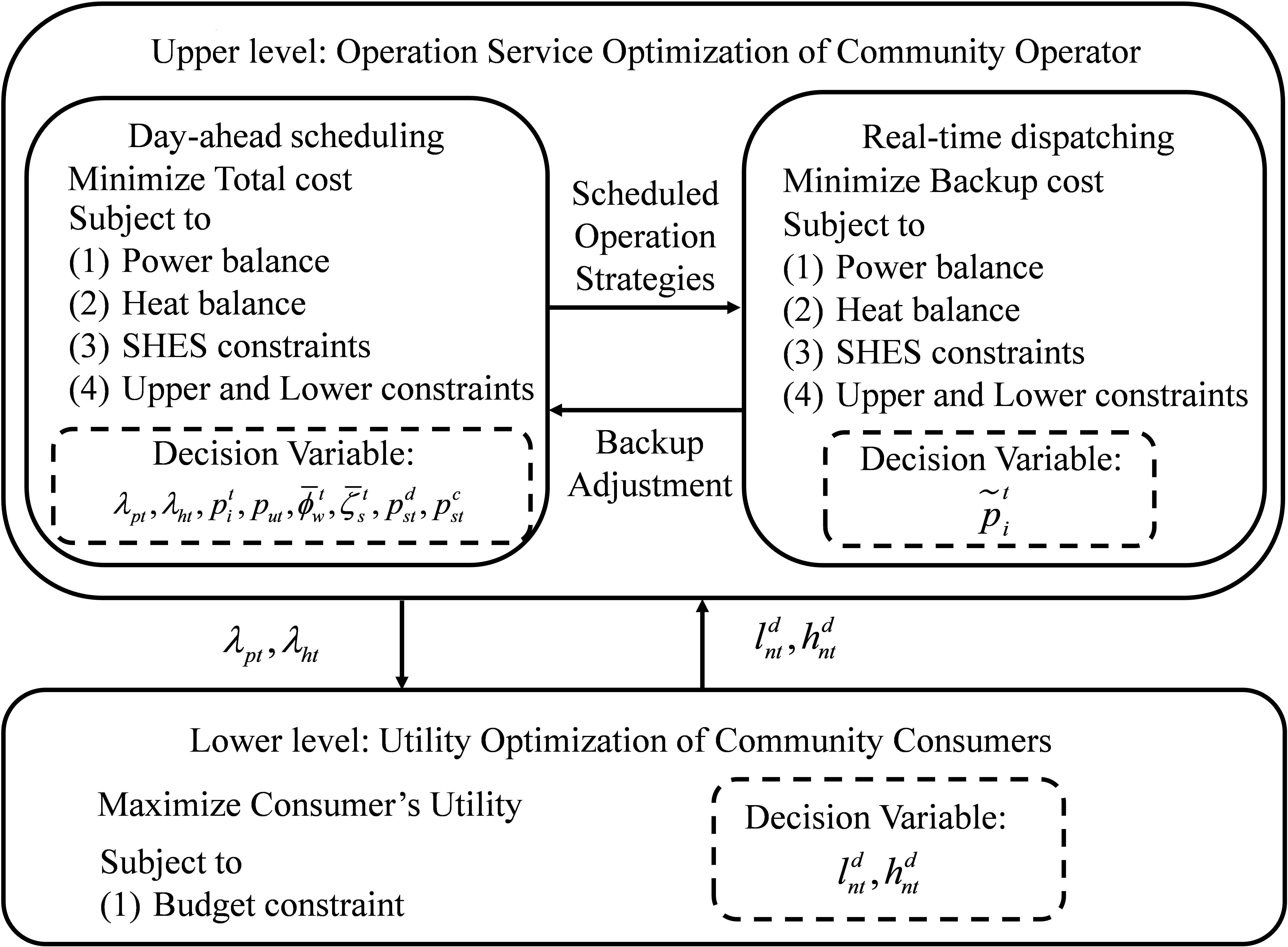}
	\caption{The solution structure of the proposed bi-level IHP model.}
	\label{Model}
\end{figure}

\subsection{Fixed-Point Algorithm of the Market Equilibrium Model}
For each generator $i \in I$, its approach involves selecting a strategy ${u_i} = \left\{ {\lambda _{pt},\lambda _{ht},p_{{i}}^{t},{p_{ut}},{\bar \phi _w^t},{\bar \zeta _s^t},p_{st}^d,p_{st}^c} \right\}$, with an objective function ${f}\left( {{u_i}} \right)$, whose constraint is ${V}\left( {{u_{ - i}},s} \right)$. For each consumer $n \in N$, its approach involves selecting a strategy ${s_n} = \left\{ {l_{nt}^d,h_{nt}^d} \right\}$, with an objective function ${g_n}\left( {{s_n}} \right)$, whose constraint is ${S_n}\left( {{s_{ - n}},u} \right)$.

\begin{myDef}
Considering the IHP system model comprising of $I$ generators and $N$ consumers, their interactions are represented as a Stackelberg game, where the community operator is a leader, endeavoring to minimize the overall expenditure  ${f_i}\left( {{u_i}} \right)$. The consumers are followers, aiming to maximize their utility ${g_n}\left( {{s_n}} \right)$, by opting for the strategies denoted by ${u_i}$ and ${s_n}$. In a formal sense, strategy profile is an equilibrium if and only if:
	\begin{equation}
		\begin{array}{l}
			u_i^* = \arg \max \left\{ {{f_i}\left( {u_m^{}} \right),\forall {u_i} \in {U_i}\left( {u_{ - i}^*,{s^*}} \right)} \right\},\quad \forall i \in I,  \\
			s_n^* = \arg \max \left\{ {{g_n}\left( {s_n^{}} \right),\forall {s_n} \in {S_n}\left( {u_{}^*,s_{ - n}^*} \right)} \right\},\quad \forall n \in N. 
		\end{array} \label{eq43}
	\end{equation}
\end{myDef}

We propose a fixed-point algorithm to find the market equilibrium. This approach performs iterative clearing of the lower-level market with fixed demands and the upper-layer market with fixed energy prices. The specific steps are outlined in Algorithm 1.

\begin{algorithm}
	%\textsl{}\setstretch{1.8}
	\renewcommand{\algorithmicrequire}{\textbf{Input:}}
	\renewcommand{\algorithmicensure}{\textbf{Output:}}
	\caption{Fixed-point algorithm}
	\label{alg1}
	\begin{algorithmic}[1]
		\STATE Initialization: Select the initial value $l_{nt,1}^d,h_{nt,1}^d$.  Set $k=1$, and set the convergence criterion parameter $\varepsilon $.
		\REPEAT
		\STATE  Solve Lower level model for Eq.(\ref{eq99}), and then obtain $\lambda _{pt,k},\lambda _{ht,k}$ . 
		\STATE $k \leftarrow k + 1$
		\STATE Update $l_{nt,k}^d,h_{nt,k}^d $ based on Equation~(\ref{eq17})
		\UNTIL $\left\| {l_{nt,k + 1}^d - l_{nt,k }^d} \right\| + \left\| {h_{nt,k + 1}^d - h_{nt,k}^d} \right\| < \varepsilon$   
		\STATE   Update $l_{nt,k}^d,h_{nt,k}^d$ based on Equation~(\ref{eq17}) 
		\ENSURE  decomposed modes $l_{nt,k}^d,h_{nt,k}^d,\lambda _{nt,k}^p,\lambda _{nt,k}^h$
	\end{algorithmic}  
\end{algorithm}

\subsection{Existence and Uniqueness of Load Equilibrium}

In this section, we discuss the existence and uniqueness of load equilibrium. 

We consider power plants. Based on the objective function (\ref{eq99}), if we represent the electricity price $\lambda_{pt}$ as a function of $l_{nt}^d$, it would exhibit an increasing affine relationship with a slope denoted by ${k}$. 

Conversely, according to (\ref{eq17}), the elastic demands decline as $\lambda_{pt}$ increases. Both relationships demonstrate continuity. Hence, an intersection must exist, implying the existence of a unique fixed-point equilibrium. Thus, The convergence of Algorithm 1 holds.

\subsection{Relaxation of Infeasible Constraints}
Substituting ${\widetilde p_i^t}$ from  (\ref{eq21}) into chance constraints  (\ref{eq22})-(\ref{eq25}), we obtain the following chance constraints:

\begin{equation}
\frac{{p_i^t - {{\overline P }_i}}}{{{\gamma _i}}} + \phi _{w,sch}^t + \zeta _{s,sch}^t \leqslant Q\left( {\widetilde \phi _w^t + \widetilde \zeta _s^t|{\alpha _{up}}} \right),\label{eq44}
\end{equation}
\begin{equation}
\frac{{p_i^t - {{\underline P }_i}}}{{{\gamma _i}}} + \phi _{w,sch}^t + \zeta _{s,sch}^t \geqslant Q\left( {\widetilde \phi _w^t + \widetilde \zeta _s^t|1 - {\alpha _{dr}}} \right),\label{eq45}
\end{equation}
\begin{equation}
\begin{gathered}
{\overline P ^L} - \sum\limits_{i = 1}^{{N_G}} {F_i^Lp_i^t}  - \sum\limits_{i = 1}^{{N_G}} {F_i^L{\gamma _i}\left( {\phi _{w,sch}^t + \zeta _{s,sch}^t} \right)}  - \sum\limits_{n = 1}^N {F_n^L{D_{nt}}}  \hfill \\
\geqslant Q\left( {F_w^L\widetilde \phi _w^t + F_s^L\widetilde \zeta _s^t - \sum\limits_{i = 1}^{{N_G}} {F_i^L{\gamma _i}\widetilde \phi _w^t}  - \sum\limits_{i = 1}^{{N_G}} {F_i^L{\gamma _i}\widetilde \zeta _s^t} |1 - {\alpha ^{L + }}} \right) \hfill \\ 
\end{gathered},  \label{eq46}
\end{equation}
\begin{equation}
\begin{gathered}
{\overline P ^L} + \sum\limits_{i = 1}^{{N_G}} {F_i^Lp_i^t}  + \sum\limits_{i = 1}^{{N_G}} {F_i^L{\gamma _i}\left( {\phi _{w,sch}^t + \zeta _{s,sch}^t} \right)}  + \sum\limits_{n = 1}^N {F_n^L{D_{nt}}}  \hfill \\
\geqslant Q\left( {  \sum\limits_{i = 1}^{{N_G}} {F_i^L{\gamma _i}\widetilde \zeta _s^t}  + \sum\limits_{i = 1}^{{N_G}} {F_i^L{\gamma _i}\widetilde \phi _w^t} - F_w^L\widetilde \phi _w^t - F_s^L\widetilde \zeta _s^t |1 - {\alpha ^{L - }}} \right) \hfill \\ 
\end{gathered} .  \label{eq47}
\end{equation}

In the DAS, considering curtailment of wind and solar generation, it is possible that the DAS is infeasible due to the no curtailment assumption. Due to the consideration of curtailment, the probability distributions of $\widetilde \phi _w^t$ and $\widetilde \zeta _s^t$ are different from $ \widetilde \phi _{w,av}^t$ and $\widetilde \zeta _{s,av}^t$.

\begin{theorem}
The following inequalities hold if ${c_j} \geqslant 0$.
\begin{equation}
\begin{gathered}
Q\left( {{c_1}\widetilde \phi _{w,av}^t + {c_2}\widetilde \zeta _s^t|p} \right) \leqslant Q\left( {{c_1}\widetilde \phi _{w,av}^t + {c_2}\widetilde \zeta _{s,av}^t|p} \right), \hfill \\
Q\left( {{c_1}\widetilde \phi _w^t + {c_2}\widetilde \zeta _{s,av}^t|p} \right) \leqslant Q\left( {{c_1}\widetilde \phi _{w,av}^t + {c_2}\widetilde \zeta _{s,av}^t|p} \right), \hfill \\ 
\end{gathered}  \label{eq49}
\end{equation}
where ${c_1}$ and ${c_2}$ are constant coefficients.
\end{theorem}

The proof of Theorem 1	can be found in Appendix B. Under Theorem 1, the curtailment of wind and solar generation will lower the p-quantile of the random variable in (\ref{eq44}) and (\ref{eq45}). Analogously, the transmission constraint (\ref{eq46}) can be relieved if the equivalent PTDF of wind and solar generations are positive. Constraint (\ref{eq47}) can be relieved by the renewable curtailment if the equivalent PTDF of wind and solar generations are negative.

By introducing slack variables, we then relax the constraints of the DAS model P1 to recognize the infeasible constraints and obtain problem P3. The objective of problem P3 is to minimize the total weighted cost of slack variables. 
\begin{equation}
\textbf{P3:}\mathop {\min }\limits_{dr{s^t},ts_{L + }^t,ts_{L - }^t} \sum\limits_{t = 1}^T {\left( {W_{Dr}^tdr{s^t} + \sum\limits_{L = 1}^{{N_L}} {W_L^t\left( {ts_{L + }^t + ts_{L - }^t} \right)} } \right)} ,\label{eq50}
\end{equation}
\begin{equation}
s.t.{\text{ }}(\ref{eq1})-(\ref{eq14}),\quad (\ref{eq21}),\quad (\ref{eq26})-(\ref{eq39}),
\end{equation}
\begin{equation}
\frac{{p_i^t - {{\overline P }_i}}}{{{\gamma _i}}} + \phi _{w,sch}^t + \zeta _{s,sch}^t \leqslant Q\left( {\widetilde \phi _w^t + \widetilde \zeta _s^t|{\alpha _{up}}} \right),\label{eq51}
\end{equation}
\begin{equation}
\frac{{p_i^t - {{\underline P }_i}}}{{{\gamma _i}}} + \phi _{w,sch}^t + \zeta _{s,sch}^t \geqslant Q\left( {\widetilde \phi _w^t + \widetilde \zeta _s^t|1 - {\alpha _{dr}}} \right) - dr{s^t},\label{eq52}
\end{equation}
\begin{equation}
\begin{gathered}
{\overline P ^L} - \sum\limits_{i = 1}^{{N_G}} {F_i^Lp_i^t}  - \sum\limits_{i = 1}^{{N_G}} {F_i^L{\gamma _i}\left( {\phi _{w,sch}^t + \zeta _{s,sch}^t} \right)}  - \sum\limits_{n = 1}^N {F_n^L{D_{nt}}}  \hfill \\
\geqslant Q\left( {F_w^L\widetilde \phi _w^t + F_s^L\widetilde \zeta _s^t - \sum\limits_{i = 1}^{{N_G}} {F_i^L{\gamma _i}\widetilde \phi _w^t}  - \sum\limits_{i = 1}^{{N_G}} {F_i^L{\gamma _i}\widetilde \zeta _s^t} |1 - {\alpha ^{L + }}} \right) - ts_{L + }^t \hfill \\ 
\end{gathered}, \label{eq53} 
\end{equation}
\begin{equation}
\begin{gathered}
{\overline P ^L} + \sum\limits_{i = 1}^{{N_G}} {F_i^Lp_i^t}  + \sum\limits_{i = 1}^{{N_G}} {F_i^L{\gamma _i}\left( {\phi _{w,sch}^t + \zeta _{s,sch}^t} \right)}  + \sum\limits_{n = 1}^N {F_n^L{D_{nt}}}  \hfill \\
\geqslant Q\left( {  \sum\limits_{i = 1}^{{N_G}} {F_i^L{\gamma _i}\widetilde \zeta _s^t}  + \sum\limits_{i = 1}^{{N_G}} {F_i^L{\gamma _i}\widetilde \phi _w^t} - F_w^L\widetilde \phi _w^t - F_s^L\widetilde \zeta _s^t |1 - {\alpha ^{L - }}} \right) - ts_{L - }^t \hfill \\ 
\end{gathered} ,  \label{eq54}
\end{equation}
\begin{equation}
dr{s^t},ts_{L + }^t,ts_{L - }^t \geqslant 0,\label{eq55}
\end{equation}
where constraints (\ref{eq51})-(\ref{eq54}) are derived from (\ref{eq44})-(\ref{eq47}) by introducing slack variables as relaxation.

\subsection{Schedule of Curtailment and Generation}
We can obtain the solution of the slack variables from problem P3. If all the slack variables are zero, the DAS model P1 has a feasible solution without curtailment. If the slack variables are nonzero, then it is necessary to set curtailment of wind and solar generation to make the DAS model P1 feasible. Thus, the following inequalities (\ref{eq56}), (\ref{eq57}) and (\ref{eq58}) hold:
\begin{equation}
Q\left( {\widetilde \phi _w^t + \widetilde \zeta _s^t|1 - {\alpha _{dr}}} \right) \leqslant Q\left( {\widetilde \phi _{w,av}^t + \widetilde \zeta _{s,av}^t|1 - {\alpha _{dr}}} \right) - dr{s^t},\label{eq56}
\end{equation}
\begin{equation}
\begin{gathered}
Q\left( {F_w^L\widetilde \phi _w^t + F_s^L\widetilde \zeta _s^t - \sum\limits_{i = 1}^{{N_G}} {F_i^L\widetilde \zeta _s^t}  - \sum\limits_{i = 1}^{{N_G}} {F_i^L\widetilde \phi _w^t} |1 - {\alpha ^{L + }}} \right) \leqslant  \hfill \\
Q\left( {\left( {F_w^L - \sum\limits_{i = 1}^{{N_G}} {F_i^L} } \right)\widetilde \phi _{w,av}^t + \left( {F_s^L - \sum\limits_{i = 1}^{{N_G}} {F_i^L} } \right)\widetilde \zeta _{s,av}^t|1 - {\alpha ^{L + }}} \right) - ts_{L + }^t, \hfill \\ 
\end{gathered}    \label{eq57}
\end{equation}
\begin{equation}
\begin{gathered}
Q\left( {\sum\limits_{i = 1}^{{N_G}} {F_i^L\widetilde \phi _w^t}  + \sum\limits_{i = 1}^{{N_G}} {F_i^L\widetilde \zeta _s^t}  - F_w^L\widetilde \phi _w^t - F_s^L\widetilde \zeta _s^t|1 - {\alpha ^{L - }}} \right) \leqslant  \hfill \\
Q\left( {\left( {\sum\limits_{i = 1}^{{N_G}} {F_i^L}  - F_w^L} \right)\widetilde \phi _{w,av}^t + \left( {\sum\limits_{i = 1}^{{N_G}} {F_i^L}  - F_s^L} \right)\widetilde \zeta _{s,av}^t|1 - {\alpha ^{L - }}} \right) - ts_{L - }^t \hfill \\ 
\end{gathered}. \label{eq58}
\end{equation}

The constraints (\ref{eq56})-(\ref{eq58}) can be reformulated in a generic format as shown below:
\begin{equation}
Q\left( {\sum\limits_{j \in \Omega } {{c_j}} \tilde r_j^t + \sum\limits_{k \notin \Omega } {{c_k}} \tilde r_{k,av}^t\mid p} \right) \leqslant a, \label{eq59}
\end{equation}
where $\tilde r_{}^t = \left\{ {\widetilde \phi _w^t,\widetilde \zeta _s^t} \right\}$, $\tilde r_{av}^t = \left\{ {\widetilde \phi _{w,av}^t,\widetilde \zeta _{s,av}^t} \right\}$, and $\Omega $ is the subscript set of wind and solar generation curtailment.  

The objective of curtailment scheduling is to minimize the overall penalty:
\begin{equation}
{\min _{\bar r_j^t(j \in \Omega )}}\sum\limits_{j \in \Omega } W_j^t\left( {\bar r_j^t} \right) \label{eq60},
\end{equation}
where $\overline r _{}^t = \left\{ {\overline \phi  _w^t,\overline \zeta  _s^t} \right\}$ and $j$ is the subscript.

Since the penalty function is nonlinear, a first-order Taylor expansion is proposed for linear approximation to obtain the curtailment scheduling model, and the curtailment penalty is a convex function according to (\ref{eq62}). 
\begin{equation}
\frac{{\partial W_j^t\left( {\bar r_j^t} \right)}}{{\partial \bar r_j^t}} = {K_j}\left[ {CDF_j^t\left( {\bar r_j^t} \right) - CDF_j^t\left( {{C_j}} \right)} \right] \leqslant 0,\label{eq61}
\end{equation}
\begin{equation}
\frac{{{\partial ^2}W_j^t\left( {\bar r_j^t} \right)}}{{\partial \bar r_j^{t2}}} = {K_j}PDF_j^t\left( {\bar r_j^t} \right) \geqslant 0.\label{eq62}
\end{equation}

\begin{theorem}
The following inequality is a sufficient condition of inequality (\ref{eq59}),  if ${c_j} \geqslant 0$ holds. 
\begin{equation}
\sum\limits_{j \in \Pi } {{c_j}} \bar r_j^t + Q\left( {{\sum _{j \in \Omega \backslash \Pi }}{c_j}\tilde r_{j,av}^t + \sum\limits_{k \notin \Omega } {{c_k}} \tilde r_{k,av}^t\mid p} \right) \leqslant a,\label{eq65}
\end{equation}
where $\Pi$ is an arbitrary subset of $\Omega $.
\end{theorem}

The proof of Theorem 2 is provided in Appendix B. Theorem 2 gives a sufficient condition of the wind and solar generation curtailments. Constraint (\ref{eq59}) can be transformed as (\ref{eq66}),  and the curtailment scheduling model can be transformed into the problem P4 as follows:
\begin{equation}
\textbf{P4:}{\min _{\bar r_j^t(j \in \Omega )}}\sum\limits_{j \in \Omega } {l_j^t} {\text{ }}\label{eq63}
\end{equation}
\begin{equation}
s.t.{\text{ }}W_j^t\left( {{r_k}} \right) + \frac{{\partial W_j^t\left( {{r_k}} \right)}}{{\partial {r_k}}}\left( {\bar r_j^t - {r_k}} \right) \leqslant l_j^t, (k = 1,2, \ldots ,M),\label{eq64}
\end{equation}
\begin{equation}
\bigcup\limits_{{\Pi _i}} {\left\{ {\sum\limits_{j \in {\Pi _i}} {{c_j}} \bar r_j^t + Q\left( {{\sum _{j \in \Omega \backslash {\Pi _i}}}{c_j}\tilde r_{j,av}^t + \sum\limits_{k \notin \Omega } {{c_k}} \tilde r_{k,av}^t\mid p} \right) \leqslant a} \right\}} .\label{eq66}
\end{equation}

The curtailment schedule can be obtained by solving problem P4. Then constraints (\ref{eq44})–(\ref{eq47}) are turned into deterministic linear constraints. The generation schedules in DAS model P1 is equivalent to the following problem P5:
\begin{equation}
\textbf{P5:}\mathop {\min }\limits_{p_i^t,\phi _{w,sch}^t,\zeta _{s,sch}^t} \sum\limits_{t = 1}^T {\sum\limits_{i = 1}^{{N_G}} {\left( {{a_i}{{\left( {p_{_i}^t} \right)}^2} + {b_i}p_{_i}^t + {c_i}} \right)} }  + \sum\limits_{t = 1}^T {{h_{gas,t}}\lambda _h^t} , \label{eq67}
\end{equation}
\begin{equation}
s.t.{\text{ }}0 \leqslant \phi _{w,sch}^t \leqslant \bar \phi _w^t,\quad 0 \leqslant \zeta _{s,sch}^t \leqslant \bar \zeta _s^t,\label{eq68}
\end{equation}
\begin{equation}
(\ref{eq1})-(\ref{eq14}),\quad (\ref{eq21}), \quad(\ref{eq26})-(\ref{eq39}),
\end{equation}
where constraint (\ref{eq68}) stipulates the upper boundary of scheduled wind and solar generations.

\section{Simulation}
In this section, a modified IEEE 33-bus power distribution system and a 32-bus district heating system are developed to evaluate and validate the effectiveness of the proposed coordinated scheduling solution for multi-timescale heat and electricity, as depicted in Fig. 4. The test system consists of generators, power-heat units, consumers and SHES. 

\begin{figure}[!t]
	\centering
	\includegraphics[width=3in]{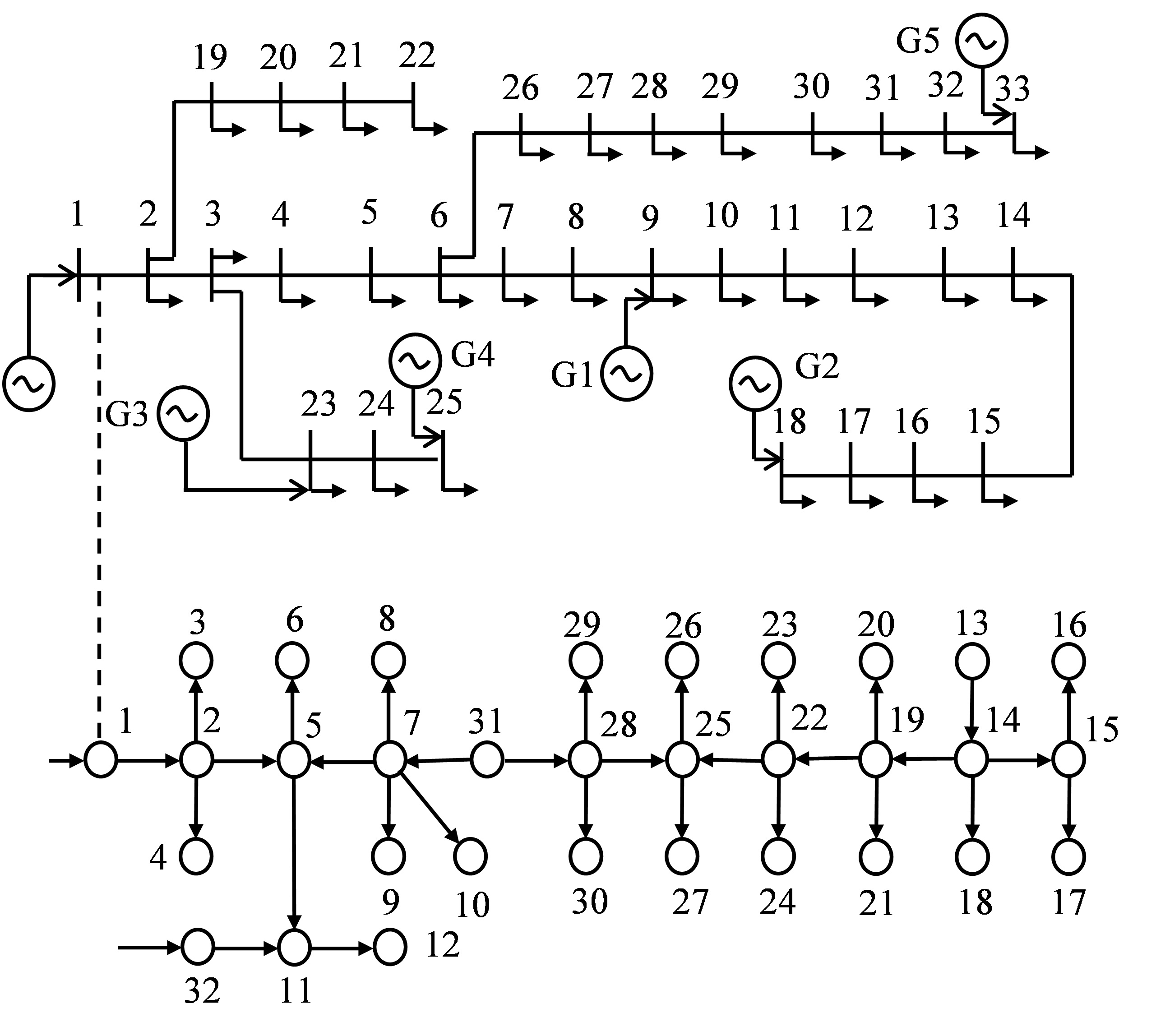}
	\caption{Schematic illustration of the 33-node power system and the 32-node heat system.}
	\label{33node}
\end{figure}

\subsection{Data}
We collected load data from the residential community in Ireland in 2009, encompassing load profiles of 200 consumers over a 24-hour period. Probability distributions for wind and photovoltaic power generation are described by GMM. The community comprises 200 consumers, a SES, and a SHS. The time cycle is 24 hours, and the time interval for the DAS of the IPH system is 1 hour. The generation data curve of the wind power in a single day assumes that the supply is stable in 1 hour. The solar and wind power generation data curve for a 24-hour period assumes a stable supply within each one-hour interval. The capacity of the transformer is 1500 kWh. The fixed power and heat load of the users is given by the actual data in Ireland in 2019 \cite{Irish}\cite{Ireland}.

For the TLOU price, the level-of-use coefficient ${k}=0.01$. As for the shared electricity storage, the charging efficiency is 90\%, and the discharging efficiency is 90\%. The capacity is 1000kWh. The maximum charging power parameter is 0.125, and the maximum discharging power parameter is 0.125. The lower boundary of the SOC is 0.1, and the upper boundary of the SOC is 0.9. The initial state is 0.2$Q_s$. As for the shared heat storage, the charging efficiency is 90\%, and the discharging efficiency is 90\%. The capacity is 1000kWh. The maximum charging power parameter is 0.125, and the maximum discharging power parameter is 0.125. The lower boundary of the SOC is 0.1, and the upper boundary of the SOC  is 0.9. The initial state is 0.2$Q_h$. The upper boundary of the flexible load is $\bar I$, and the parameter $\alpha$ is set as 0.3. 

The experiments are conducted on a laptop equipped with an Intel Core i7-7500U CPU running at 2.7GHz and 8GB of RAM.  The simulation platform is developed using MATLAB, YALMIP, and Gurobi, where the equilibrium problem is solved in MATLAB using the YALMIP interface, while the commercial solver Gurobi is employed to solve the MILP.

\subsection{Benchmark Case}

The DAS of the SOC of the shared electricity and heat storage is shown in Fig. 5. The upper graph corresponds to the SOC of the shared electricity storage, where the blue line represents the SOC of the SES, the red line represents the charging power, and the purple line represents the discharging power, respectively. The left axis represents the SOC of electricity storage, and the right axis represents the charging and discharging power. We can see that the shared electricity storage discharged from 1 a.m. to 7 a.m., followed by charging from 7 a.m. to 6 p.m. Subsequently, it discharged again from 6 p.m. to 10 p.m., and then charged at 11 p.m., ultimately discharging at 12 p.m.

In Fig. 5, the lower graph represents the SOC of the SHS, where the blue line represents the SOC of the shared heat storage, the red line represents the charging heat, and the purple line represents the discharging heat, respectively. The left axis represents the SOC of the SHS, and the right axis represents the charging and discharging heat. We can see that the SHS charged from 2 a.m. to 6 a.m. and discharged from 7 p.m. to 24 p.m.

\begin{figure}[!t]
	\centering
	\includegraphics[width=3.2in]{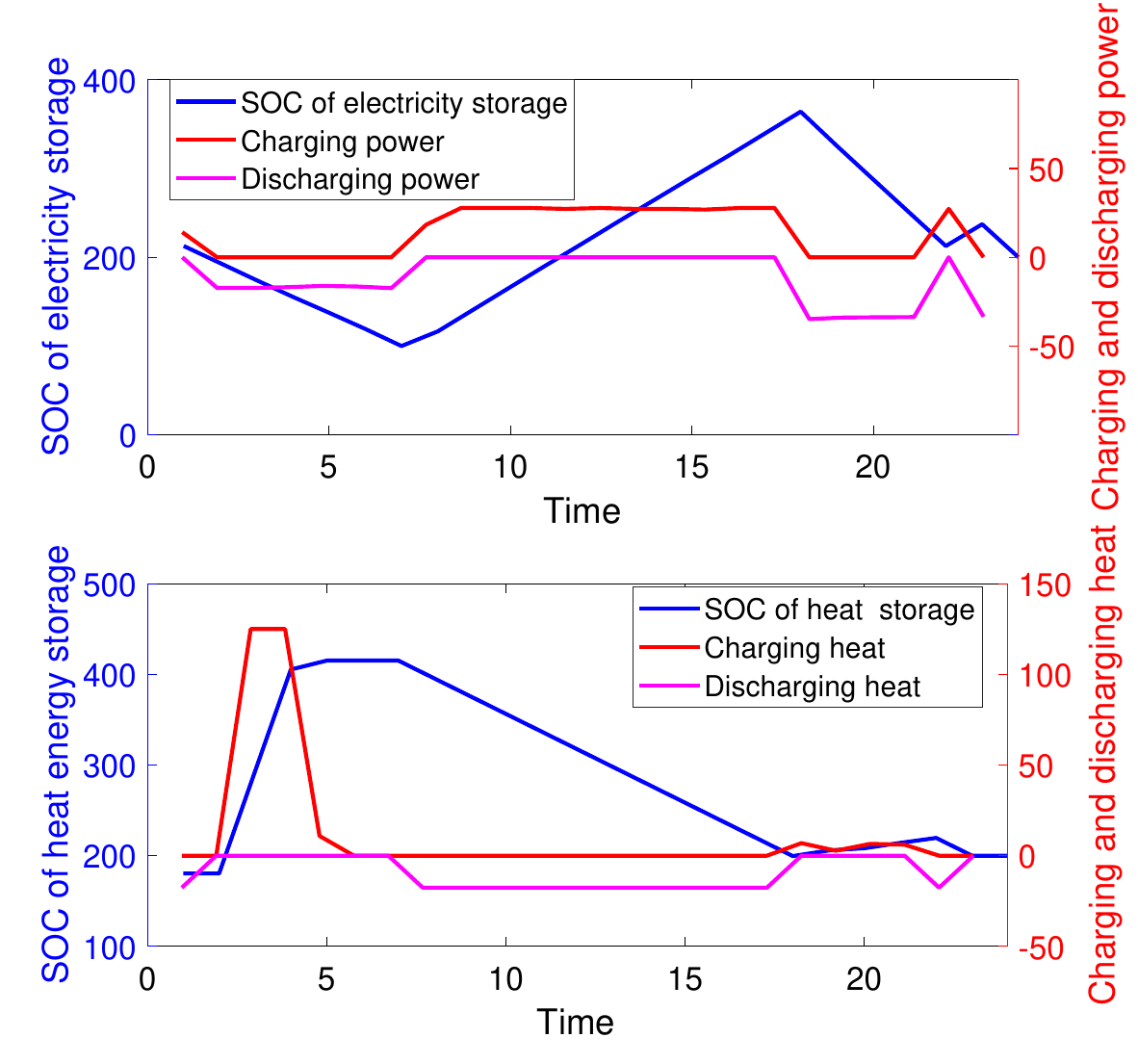}
	\caption{Day-ahead scheduling of SOC of the shared electricity and heat storage in the IEEE 33-bus power distribution system and the 32-bus district heating system.}
	\label{soces}
\end{figure}

The DAS of power/heat demands and the power/heat prices are shown in Fig. 6, where the blue line represents power/heat demands, and the red line represents the power/heat price, respectively. The left axis represents the power/heat demands, and the right axis represents the real-time power/heat price. We can see that power demands are high when real time electricity price is low, the heat demands are high when the real time gas price is high, because the heat demands can be supplied by CHP units.

\begin{figure}[!t]
	\centering
	\includegraphics[width=3.2in]{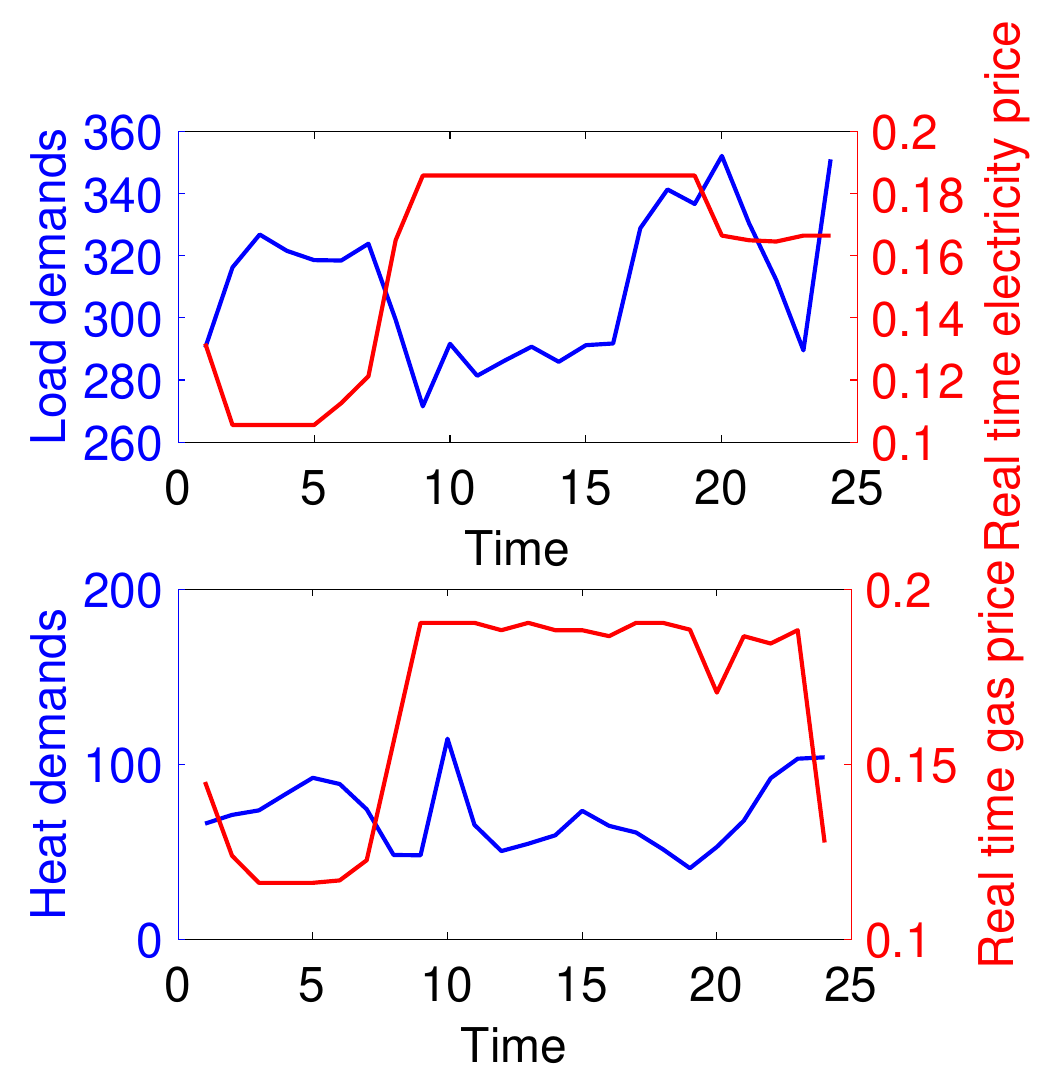}
	\caption{Day-ahead scheduling of power/heat demands and power/heat price.}
	\label{loadprice}
\end{figure}

RTD of heat and power utilization is shown in Fig. 7. Heat utilization includes gas, P2H, discharging and charging heat of SHS. Power utilization includes demands, P2H, charging and discharging power of SES. During 2:00-5:00, SHS continues charging when the real-time gas price is low. During 8:00-23:00, SHS continues discharging when the real-time gas price is high. During 8:00-18:00, SES continues charging when the wind and solar generations are high. During 19:00-22:00, SES continues discharging when the wind and solar generations are low.

\begin{figure}[!t]
	\centering
	\includegraphics[width=3.2in]{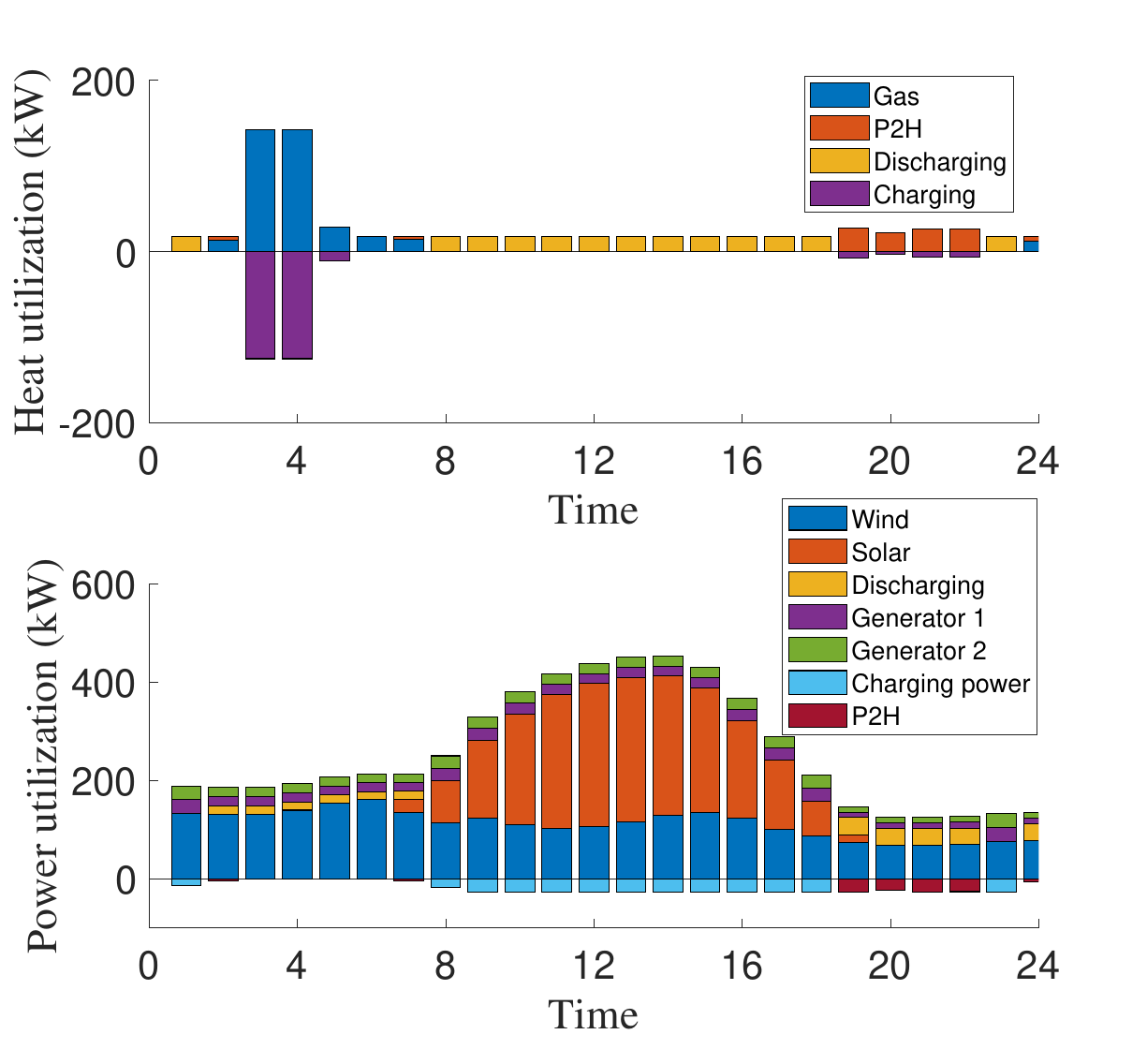}
	\caption{Real-time dispatching of heat and power utilization.}
	\label{Day}
\end{figure}

To analyze the influence of power transformed to heat $p_{ut}$, the power to heat value describes the power to heat power with the variation of time. The power to heat turbine works from 7 p.m. to 10 p.m. The optimization problem yields an objective value of $392.73$. However, when excluding $p_{ut}$, the objective value increases to $407.12$, highlighting the significant impact of $p_{ut}$ on the overall cost.

\subsection{Sensitivity Analysis on the Capacity of SHES}

We test the SHES capacity from 0 to 2000kWh to investigate the sensitivity analysis of the capacity of SHES on the operation cost in a day. The SHES  is expected to operate for ten years and the investment cost of the shared energy storage is 300 euros/kWh \cite{Zhang2020}. Results are shown in Fig. 8. The operation cost decreases with the increase of the capacity of SHES at first. However, when the capacity of SHES reaches 300 KWh, the total cost starts to increase with the capacity of SHES. 
\begin{figure}[!t]
	\centering
	\includegraphics[width=3.2in]{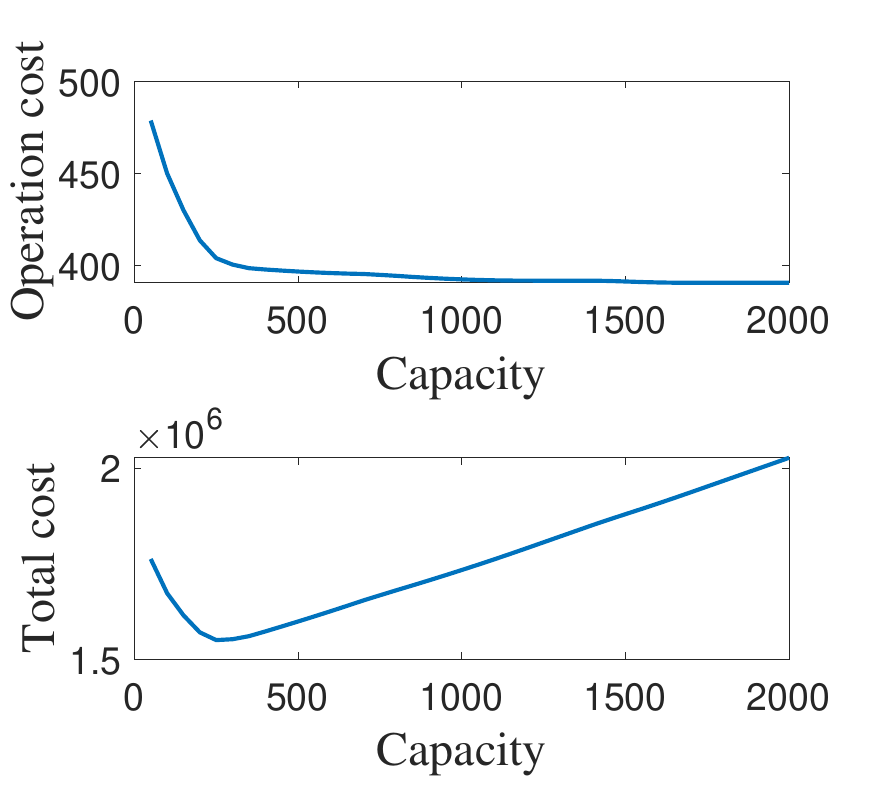}
	\caption{Sensitivity analysis of the operation cost and total cost on the capacity of SHES.}
	\label{capacity}
\end{figure}

\subsection{Sensitivity Analysis on Utility Function Parameters}
The impact of the utility function parameter  $\alpha$ on the power and heat load is investigated. We change the parameter $\alpha$  from $0.1$ to $0.9$ and solve the problem. Results are shown in Table II. It is observed that the power load increases with the growth of the utility function parameter and that the heat load decreases with the growth of the utility function parameter. 

\begin{table}[ht]
	\renewcommand{\arraystretch}{1.3}
	\renewcommand{\tabcolsep}{1em}
	\scriptsize
	\centering
	\caption{The sensitivity analysis on utility function parameters.}
	\label{tab:comparison-2}
	\begin{tabular}{ccccccc}
		\hline  {Utility function parameter} & {Total power load} & {Total heat load}\\
		\hline
		{0.1} & $4106.0$& $2014.0$  \\
		{0.3 }& $7447.2$& $1697.8$\\
		{0.5 }& $10788.0$& $1381.6$\\
		{0.7 }& $14130.0$& $1065.5$  \\
		{0.9 }& $17471.0$& $749.2$\\
		\hline
	\end{tabular}
\end{table}

\section{Conclusion}
In order to investigate the solution to the scheduling optimization problem of the multi-time-scale operation in the integrated heat-power system with SHES, we propose a Stackelberg equilibrium game model of the IHP system, where the community operator acts as a leader and the consumers act as followers. We explicitly model the interaction between the heat and power generators, SHES, and demands. We consider the generation uncertainties from renewable energy and the demand uncertainties according to the Cobb-Douglous function. To illustrate the strategies of the community operator, we propose a new chance-constrained energy dispatching strategy. In DAS, we adopt the GMM method for wind and solar power. In RTD, the uncertainties of wind and solar power are balanced by generators.  Through the KKT conditions of the consumers, we transform the optimization problem into a MILP optimization problem. We propose a fixed-point algorithm and illustrate the existence and uniqueness of the load equilibrium. Moreover, we propose a recognition method of infeasible constraints and a schedule of curtailment and generation. The DAS can be transformed into a deterministic optimization problem and be solved by linear optimization method.  Simulations verified the effectiveness of the proposed chance-constrained multi-timescale heat-power model.

\bibliographystyle{IEEEtran}

\bibliography{IEEEabrv,mybib}

\appendices
\makeatletter
\@addtoreset{equation}{section}
\@addtoreset{theorem}{section}
\makeatother
\setcounter{equation}{0}  
\renewcommand{\theequation}{A.\arabic{equation}}
\renewcommand{\thetheorem}{A.\arabic{theorem}}
\section{Cobb-Douglas Utility}
The Cobb-Douglas utility is as follows: 
\begin{equation}
U(x,y) = {x^\alpha }{y^{(1 - \alpha )}},0 < \alpha  < 1, \label{eq69}
\end{equation}
where $x$ and $y$ are products consumption, and $\alpha $ is a preference coefficient. The Cobb-Douglas function has the following properties.

\textit{(1) Positive Marginal Utility:} In the Cobb-Douglas function, we can derive the marginal utilities of the products $x$ and $y$ are positive in (\ref{eq70}).
\begin{equation}
\frac{\partial U}{\partial x}=\alpha x^{\alpha-1} y^{1-\alpha}, \frac{\partial U}{\partial y}=(1-\alpha) x^\alpha y^{-\alpha}, \label{eq70}
\end{equation}

\textit{(2) Diminishing Marginal Utility:} In the Cobb-Douglas function, we can derive the second derivatives of products $x$ and $y$ are both negative.
\begin{equation}
\frac{\partial^2 U}{\partial x^2}=\alpha(\alpha-1) x^{\alpha-2} y^{1-\alpha}, \frac{\partial^2 U}{\partial y^2}=\alpha(\alpha-1) x^\alpha y^{-\alpha-1},\label{eq71}
\end{equation}

\textit{(3) Concave Marginal Utility:} In the Cobb-Douglas function, we can derive the Hessian matrix of products $x$ and $y$ is  negative semidefinite, therefore, $U(x,y)$ is concave.
\begin{equation}
\left[\begin{array}{ll}
\frac{\partial^2 U}{\partial x^2} & \frac{\partial^2 U}{\partial x \partial y} \\
\frac{\partial^2 U}{\partial y \partial x} & \frac{\partial^2 U}{\partial y^2}
\end{array}\right]=\left[\begin{array}{cc}
\alpha(\alpha-1) x^{\alpha-2} y^{1-\alpha} & \alpha(1-\alpha) x^{\alpha-1} y^{-\alpha} \\
\alpha(1-\alpha) x^{\alpha-1} y^{-\alpha} & \alpha(\alpha-1) x^\alpha y^{-1-\alpha}
\end{array}\right],\label{eq72}
\end{equation}
\begin{equation}
\frac{\partial^2 U}{\partial x^2}<0, \frac{\partial^2 U}{\partial y^2}<0, \frac{\partial^2 U}{\partial x^2} \frac{\partial^2 U}{\partial y^2}-\frac{\partial^2 U}{\partial x \partial y} \frac{\partial^2 U}{\partial y \partial x}=0,\label{eq73}
\end{equation}

The KKT conditions of Eqs. (\ref{eq15})-(\ref{eq16}) are given as follows.
\begin{equation}
\alpha {\left( {\frac{{{h_{nt}^d}}}{{{l_{nt}^d}}}} \right)^{1 - \alpha }} - \xi \lambda _{pt} = 0,\label{eq74}
\end{equation}
\begin{equation}
(1 - \alpha ){\left( {\frac{{{l_{nt}^d}}}{{{h_{nt}^d}}}} \right)^\alpha } - \xi \lambda _{ht} = 0,\label{eq75}
\end{equation}
\begin{equation}
0 \le \xi  \bot \left( {\bar O - \lambda _d^p{l_{nt}^d} - \lambda _d^h{h_{nt}^d}} \right) \ge 0.\label{eq76}
\end{equation}

Based on the constraint (\ref{eq76}), we can deduce that $\xi  \ne 0$, indicating that ${\lambda _d^pl_{nt}^d + \lambda _d^hh_{nt}^d \le \bar O}$ represents the optimal solution. Subsequently, by evaluating ${\rm{(\ref{eq74})}} \times {l_{nt}^d} + (\ref{eq75}) \times {h_{nt}^d}$, we obtain
\begin{equation}
{\left( {{l_{nt}^d}} \right)^\alpha }{\left( {{h_{nt}^d}} \right)^{1 - \alpha }} = \xi \left( {\lambda _{pt}{l_{nt}^d} + \lambda _{ht}{h_{nt}^d}} \right) = \xi \bar O.\label{eq77}
\end{equation}

Substitute (\ref{eq77}) into ${\rm{(\ref{eq74})}}$ and $(\ref{eq75})$, we obtain
\begin{equation}
\frac{{\lambda _{pt}{l_{nt}^d}}}{\alpha } = \frac{{\lambda _{ht}{h_{nt}^d}}}{{1 - \alpha }} = \bar O.\label{eq78}
\end{equation}

\setcounter{equation}{0}  
\renewcommand{\theequation}{B.\arabic{equation}}
\renewcommand{\thetheorem}{B.\arabic{theorem}}
\section{Proof Of Theorem 1 and 2}
If $\widetilde \delta  \leqslant \widetilde \gamma $ holds, then $\Pr \left( {\widetilde \delta \leqslant a} \right) \geqslant \Pr \left( {\widetilde \gamma \leqslant a} \right)$. According to the monotonicity of the cumulative density function, the inequality in (\ref{eq80}) holds. 
\begin{equation}
\Pr \left( {\widetilde \delta \leqslant Q\left( {\widetilde \gamma|p} \right)} \right) \geqslant \Pr \left( {\widetilde \gamma \leqslant Q\left( {\widetilde \gamma|p} \right)} \right) = \Pr \left( {\widetilde \delta \leqslant Q\left( {\widetilde \delta|p} \right)} \right),\label{eq79}
\end{equation}
\begin{equation}
Q\left( {\widetilde \delta|p} \right) \leqslant Q\left( {\widetilde \gamma|p} \right),\label{eq80}
\end{equation}

Since $\widetilde \phi _w^t \leqslant \widetilde \phi _{w,av}^t$, and $\widetilde \zeta _s^t \leqslant \widetilde \zeta _{s,av}^t$, Theorem 1 holds due to (\ref{eq80}). Since $\tilde r_j^t \leqslant \bar r_j^t$, and $\tilde r_j^t \leqslant \tilde r_{j,av}^t$, Theorem 2 can be induced according to (\ref{eq81}).
\begin{equation}
\begin{gathered}
Q\left( {\sum\limits_{j \in \Omega } {{c_j}} \tilde r_j^t + \sum\limits_{k \notin \Omega } {{c_k}} \tilde r_{k,av}^t\mid p} \right) \hfill \\
\leqslant Q\left( {\sum\limits_{j \in \Pi } {{c_j}} \bar r_j^t + \sum\limits_{j \in \Omega \backslash \Pi } {{c_j}\tilde r_{j,av}^t}  + \sum\limits_{k \notin \Omega } {{c_k}} \tilde r_{k,av}^t\mid p} \right) \hfill \\
= \sum\limits_{j \in \Pi } {{c_j}} \bar r_j^t + Q\left( {\sum\limits_{j \in \Omega \backslash \Pi } {{c_j}\tilde r_{j,av}^t}  + \sum\limits_{k \notin \Omega } {{c_k}} \tilde r_{k,av}^t\mid p} \right) \hfill \\ 
\end{gathered} . \label{eq81}
\end{equation}

\end{document}